\documentclass[article,nojss,shortnames]{jss}

\let\maxwidth\textwidth

\usepackage[]{changes}

\usepackage{thumbpdf}
\usepackage{amsfonts,amstext,amsmath,amssymb,amsthm}
\usepackage{accents}
\usepackage{xcolor}
\usepackage{rotating}
\usepackage{verbatim}
\usepackage[graphicx]{realboxes}
\usepackage{xspace}
\usepackage{lscape}
\usepackage{fmtcount}
\usepackage{needspace}

\usepackage{accents}

\theoremstyle{example}

\newcommand{\NAMI}{nonparanormal adjusted marginal inference\xspace}

\usepackage{booktabs}
\usepackage{nicefrac}
\usepackage{float}

\newcommand{\rY}{Y}
\newcommand{\rX}{\mX}

\newcommand{\ry}{y}
\newcommand{\rx}{\xvec}

\newcommand{\samY}{\mathcal{Y}}
\newcommand{\samX}{\mathcal{X}}

\newcommand{\h}{h}

\newcommand{\eparm}{\vartheta}

\newcommand{\ie}{\textit{i.e.}~}
\newcommand{\eg}{\textit{e.g.}~}

\renewcommand{\Prob}{\mathbb{P}}

\newcommand{\Var}{\mathbb{V}}
\newcommand{\RR}{\mathbb{R}}

\usepackage{dsfont}

\def \expit {\text{logit}^{-1}}
\def \cloglog {\text{cloglog}}

 \DeclareMathOperator{\ND}{N}

\def \hvec {\text{\boldmath$h$}}

\def \xvec {\text{\boldmath$x$}}    \def \mX {\text{\boldmath$X$}}
    
\def \zvec {\text{\boldmath$z$}}

\def \thetavec        {\text{\boldmath$\theta$}}

\def \lambdavec       {\text{\boldmath$\lambda$}}

\def \mSigma   {\mathbf{\Sigma}}

\def \mPhi     {\mathbf{\Phi}}

\def \mOmega   {\mathbf{\Omega}}

\def \nullvec {\mathbf{0}}

\usepackage[utf8]{inputenc}
\usepackage{enumitem}

\makeatletter
\renewcommand*\env@matrix[1][\arraystretch]{%
	\edef\arraystretch{#1}%
	\hskip -\arraycolsep
	\let\@ifnextchar\new@ifnextchar
	\array{*\c@MaxMatrixCols c}}
\makeatother

\usepackage{colortbl}
\usepackage{longtable}

\usepackage{makecell, caption}
\usepackage{subcaption}

\makeatletter
\def\maxwidth{ %
  \ifdim\Gin@nat@width>\linewidth
    \linewidth
  \else
    \Gin@nat@width
  \fi
}
\makeatother

\definecolor{fgcolor}{rgb}{0.345, 0.345, 0.345}
\newcommand{\hlnum}[1]{\textcolor[rgb]{0.686,0.059,0.569}{#1}}%
\newcommand{\hlsng}[1]{\textcolor[rgb]{0.192,0.494,0.8}{#1}}%
\newcommand{\hlcom}[1]{\textcolor[rgb]{0.678,0.584,0.686}{\textit{#1}}}%
\newcommand{\hlopt}[1]{\textcolor[rgb]{0,0,0}{#1}}%
\newcommand{\hldef}[1]{\textcolor[rgb]{0.345,0.345,0.345}{#1}}%
\newcommand{\hlkwb}[1]{\textcolor[rgb]{0.69,0.353,0.396}{#1}}%
\newcommand{\hlkwc}[1]{\textcolor[rgb]{0.333,0.667,0.333}{#1}}%
\newcommand{\hlkwd}[1]{\textcolor[rgb]{0.737,0.353,0.396}{\textbf{#1}}}%

\usepackage{framed}
\makeatletter
\newenvironment{kframe}{%
 \def\at@end@of@kframe{}%
 \ifinner\ifhmode%
  \def\at@end@of@kframe{\end{minipage}}%
  \begin{minipage}{\columnwidth}%
 \fi\fi%
 \def\FrameCommand##1{\hskip\@totalleftmargin \hskip-\fboxsep
 \colorbox{shadecolor}{##1}\hskip-\fboxsep
     \hskip-\linewidth \hskip-\@totalleftmargin \hskip\columnwidth}%
 \MakeFramed {\advance\hsize-\width
   \@totalleftmargin\z@ \linewidth\hsize
   \@setminipage}}%
 {\par\unskip\endMakeFramed%
 \at@end@of@kframe}
\makeatother

\definecolor{shadecolor}{rgb}{.97, .97, .97}
\definecolor{messagecolor}{rgb}{0, 0, 0}
\definecolor{warningcolor}{rgb}{1, 0, 1}
\definecolor{errorcolor}{rgb}{1, 0, 0}
\newenvironment{knitrout}{}{} 

\usepackage{alltt}
\IfFileExists{upquote.sty}{\usepackage{upquote}}{}

\author{Leticia Wüthrich\\Universit\"at Z\"urich
  \And Mark Baillie\\Novartis Pharma AG
  \And Susanne Dandl\\Universit\"at Z\"urich}
\Plainauthor{Wüthrich, Baillie \& Dandl}

\title{Randomization-Based \\
	     Nonparanormal Adjusted Marginal Inference}
\Plaintitle{Randomization-Based NAMI}
\Shorttitle{Randomization-Based NAMI}

\Abstract{

Adjusting for prognostic baseline covariates in the analysis of randomized controlled 
trials is recommended to increase the precision of treatment effect estimates. Such 
adjustment, however, introduces a complication known as noncollapsibility: for effect 
measures such as the log-odds or log-hazard ratio, the estimated treatment effect 
becomes conditional instead of marginal. Methods that target a marginal estimand 
even after covariate adjustment have been developed, but typically rely on population-based 
inference, which assumes patients are randomly sampled from some larger population, 
rarely true in practice. Such methods have been shown to inflate type~I error, particularly 
in small samples or when many covariates are adjusted for. Inference based on randomization 
requires no such assumption, as it relies on the randomization scheme, which is known 
exactly, yielding a test that is exact even in finite samples. We present a randomization-based 
test for covariate-adjusted marginal treatment effects, built on nonparanormal adjusted 
marginal inference, applicable to continuous, binary, ordinal, and time-to-event outcomes. 
In simulation studies, this test controls type~I error where the standard Wald test 
does not. Applied to 20 historical phase~III time-to-event randomized trials, the covariate-adjusted 
estimator achieves precision gains comparable to an existing competitor method. }

\Plainkeywords{marginal effect, randomization test, noncollapsibility, covariate adjustment, randomized trial, transformation model}

\Address{
  Susanne Dandl, Leticia Wüthrich\\
  Institut f\"ur Epidemiologie, Biostatistik und Pr\"avention \\
  Universit\"at Z\"urich \\
  Hirschengraben 84, CH-8001 Z\"urich, Switzerland \\
  Email: \texttt{Susanne.Dandl@uzh.ch}
}

\begin{document}

\section{Introduction}
\label{sec:introduction}

Adjusting for prognostic baseline covariates in the primary analysis of
randomized controlled trials (RCTs) is recommended because it can increase
the precision of treatment effect estimates
\citep{food_and_drug_administration_adjusting_2023,van_lancker_covariate_2024}. 
However, directly including covariates in
the linear predictor of a regression model introduces a complication: the
treatment effect becomes conditional rather than marginal. For many widely
used effect measures, such as the log-odds ratio and the log-hazard ratio,
the two do not coincide, even in the absence of confounding, a phenomenon
known as noncollapsibility \citep{greenland_confounding_1999,daniel_making_2021}.
Because the estimand of interest in RCTs is often the marginal effect,
several methods have been developed to obtain marginal treatment effect
estimates while still benefiting from covariate adjustment
\citep{Tsiatis_Davidian_2008,Zhang_Tsiatis_2008,Lu_Tsiatis_2008,Ye_Shao_Yi_2024, Dandl_Hothorn_2026}.

Yet, the inference procedures accompanying these methods typically rely on
population-based asymptotic theory \citep{Rosenberger2019:randomization}. 
Standard errors and hypothesis
tests are derived from the sampling distribution of an estimator across
hypothetical repeated samples. Such approximations can become unreliable
when the number of estimated parameters is large relative to the sample
size, even if the assumed model is correctly specified
\citep{Portnoy1988:exponential,MacKinnon1985:hccme}. This is particularly
relevant for covariate adjustment, where additional covariates typically
introduce additional parameters. Indeed, simulation studies have reported
type I error rates above the nominal level when many covariates are adjusted
for relative to the available sample size
\citep{Tackney2023:covadj,Zhang2025:scoretest,Dandl_Hothorn_2026}.

Randomization itself offers a different basis for inference
\citep{Cox2009:randomization,Rosenberger2019:randomization,Guo2023:oaxacablinder}.
Unlike population-based inference, which treats the patients' outcomes as
random according to some assumed data-generating process, randomization-based
inference conditions on the enrolled patients and takes the treatment
assignment as the source of randomness.
This distinction is important because the mechanism governing treatment
assignment is specified by the trial design and is therefore exactly known. Under
the sharp null hypothesis of no treatment effect, each patient's outcome
would have been the same regardless of which treatment they received. The
observed outcomes could therefore equally well have arisen under any other
assignment permitted by the randomization design. Comparing the observed
test statistic with its distribution over these possible assignments
provides an exact reference distribution for testing the null hypothesis
\citep{Fisher1935:design,Kempthorne1955:randomization,Edgington2007:randomization}.
When this distribution is computed from the randomization scheme itself, the
resulting randomization test controls the type I error rate under the sharp
null without requiring a model for the outcome distribution. The
choice of test statistic, even if motivated by a parametric model, can
therefore affect power without compromising the validity of the test,
provided the randomization distribution is constructed correctly
\citep{Rosenberger2019:randomization}.

FDA guidance explicitly
acknowledges that randomization tests can be used together with covariate
adjustment, although it provides little detail on their implementation
\citep{food_and_drug_administration_adjusting_2023}.
Existing approaches either target collapsible effect measures
\citep{Rosenbaum2002:covadj,Zhao2021:frt} or derive their test statistics
from models targeting the
\textit{conditional} treatment effect for noncollapsible effect
measures \citep{Gail1988:randtest}.
Even though randomization tests of the sharp
null have no estimand of their own, and their validity does not depend
on the model behind the statistic,
that model still fixes the direction of the test, supplies the estimate
reported beside it, and can affect its power against different
alternatives \citep{Senn2004:addedvalues}.
To the best of our knowledge, no covariate-adjusted randomization test for
noncollapsible effect measures has been constructed from a model that
directly parametrizes the marginal treatment effect.

In this work, we address this gap by combining nonparanormal adjusted
marginal inference \citep[NAMI;][]{Dandl_Hothorn_2026} with a general
permutation-testing framework \citep{Hothorn2006:lego}. 
We choose NAMI because, compared to other adjustment methods
\citep[e.g.,][]{Tsiatis_Davidian_2008,Zhang_Tsiatis_2008,Lu_Tsiatis_2008,Ye_Shao_Yi_2024},
it fulfills the following: (1) it \emph{directly} estimates a marginal
treatment effect while adjusting for prognostic covariates, thereby 
resolving the issue of noncollapsibility; (2) it is fully parametric and 
likelihood-based, so the score residuals underlying the randomization 
test are readily available; and (3) it builds on transformation models, 
accommodating a wide range of treatment effect measures, outcome types, 
and censoring mechanisms. 

We show how randomization-based inference can be incorporated into NAMI by
using its score residuals to construct a test statistic within a general
permutation-testing framework \citep{Hothorn2006:lego}. The idea is to fit NAMI under the
null hypothesis and use the resulting score residuals, which target the
marginal rather than the conditional effect, to measure the association
between the outcome and the randomized treatment assignment. In this
way, the model determines the test statistic, while the randomization scheme
determines the reference distribution against which it is assessed. This
reference distribution, defined by the treatment assignments permitted by
the trial's design, can be obtained by enumerating all permitted
assignments, approximated through Monte Carlo resampling, or approximated
using the asymptotic distribution of the permutation statistic
\citep{Strasser1999:permutation}.
Evaluating the exact randomization distribution provides finite-sample type
I error control under the sharp null.

We start by describing transformation models and their multivariate
extension to NAMI in Section~\ref{sec:methods}, before presenting the
score-residual randomization test and its extensions to stratified
designs and to more than two treatment groups. We then
evaluate its finite-sample performance
in simulations in Section~\ref{sec:simulation}, showing that the test
controls the type I error rate in
settings where the population-based NAMI Wald test does not. In
Section~\ref{sec:application}, we demonstrate the practical feasibility of
the test on 20 historical phase~III clinical trials and show that its
underlying estimator achieves efficiency gains comparable to an existing
competitor method.
 
\section{Methods}
\label{sec:methods}

Consider an outcome $\rY \in \samY$ on an at least ordered sample space $\samY$,
and a binary treatment indicator $W \in \{0, 1\}$, with $W = 0$ denoting control
and $W = 1$ denoting treatment. Let $\rX = (X_1, \ldots, X_{J-1})$ denote $J - 1$
baseline covariates, with $X_j \in \samX_j$ for $j = 1, \ldots, J - 1$ each assumed
to take values in an at least ordered sample space $\samX_j$. We first consider a
simple randomization setting, so that treatment probability does not depend on
$\rX$ or $\rY$; extensions to stratified randomization and to settings with
more than two treatment groups are discussed in Sections~\ref{sec:stratification}
and~\ref{sec:moretreats}.

Let $F_0$ and $F_1$ denote the distributions of the outcome under control and
treatment, respectively, marginal with respect to the covariates $\rX$, with
$F_w(\ry) = \Prob(\rY \le \ry \mid W = w)$ for $w \in \{0, 1\}$. We are interested
in the marginal treatment effect $\tau := \tau(F_0, F_1) \in \RR$, which represents
the discrepancy between $F_0$ and $F_1$.

Our goal is to present a general permutation-based procedure for inference on the
treatment effect $\tau$ that adjusts for covariates and derives its validity from
the randomization distribution of $W$, rather than from an invoked population-based
model. Transformation models provide a flexible building block for this purpose.
Section~\ref{sec:unadjusted} reviews the univariate transformation model 
\citep{hothorn_moest_buehlmann_2017}, a unifying framework linking many classical 
models across outcome types and treatment effect measures. Section~\ref{sec:adjusted} 
presents its extension to covariate-adjusted marginal inference via NAMI
\citep{Dandl_Hothorn_2026}.
Finally, in Section~\ref{sec:likelihood_score}, we propose a randomization-based
test statistic built from the score residuals of NAMI under the null
hypothesis of no treatment effect. These score residuals can be
understood as a
covariate-adjusted quantity measuring the deviation between the observed
outcome and the outcome predicted by the covariates. Testing for a
treatment effect then amounts to assessing whether the score residuals
are associated with treatment assignment, using the randomization
distribution of $W$.

\subsection{Univariate transformation models}
\label{sec:unadjusted}

Transformation models, as described by \citet{hothorn_moest_buehlmann_2017}, describe
the entire conditional distribution of the outcome $\rY$. This allows many different 
outcome types and censoring mechanisms to be described by a common framework. Given
treatment $w \in \{0, 1\}$, we write

\begin{equation}
    F_w(\ry) = \Prob(\rY \le \ry \mid W = w) = G\bigl(\h(\ry) - \tau w\bigr),
    \label{eq:trafo-model}
\end{equation}

where $G: \RR \to [0,1]$ is an a priori chosen cumulative distribution function, $\h: \samY \to
\RR$ is an unknown monotone nondecreasing transformation function, and $\tau$ is the
marginal treatment effect, defined as a shift on the scale of $G$. Setting
$w = 0$ in~\eqref{eq:trafo-model} gives
$\h(\ry) = G^{-1}\{F_0(\ry)\}$: $\h$ maps the reference group's cumulative
distribution function onto the quantile scale of $G$. Because the functional 
form of $\h$ can be chosen to match the outcome type independently of $G$, this 
formulation separates the scale of the treatment effect, set by $G$, from the 
shape of the outcome's distribution, set by $\h$, without imposing a
distributional assumption on $\rY$ unless $\h$ itself is restricted to a
specific form.
For a discrete, ordered outcome with $K$ categories $y_1 < \cdots < y_K$, $\h$ 
is parametrized as a step function with $K - 1$ free steps. For a continuous 
outcome, $\h$ can instead be parametrized as a smooth flexible function, for 
example using a  polynomial in Bernstein form.

The treatment effect $\tau = G^{-1}\{F_0(\ry)\} - G^{-1}\{F_1(\ry)\}$ is assumed
to be constant across $\ry$, \ie it represents a location shift on the scale of 
$G$. This is equivalent to the familiar proportional odds assumption (when $G = 
\expit$) and the proportional hazards assumption (when $G = \cloglog^{-1}$).
The following three special cases of \eqref{eq:trafo-model} show how classical
models arise from different choices of $G$ and $\h$; a more comprehensive
list can be found in \citet{hothorn_moest_buehlmann_2017}.

(i) \textbf{Linear regression and Box-Cox models.} For a continuous outcome $\rY \in \RR$, a normal
distribution can be imposed by setting $G = \Phi$ (the standard normal distribution)
and constraining a linear transformation $\h(\ry) = \eparm_1 + \eparm_2 \ry$.
This is because $\rY$ can only be mapped onto the standard normal scale by a linear 
transformation if $\rY$ is already normally distributed. The model becomes 
$F_w(\ry) = \Phi(\eparm_1 + \eparm_2 \ry - \tau w)$, \ie $\rY \mid W=w \sim 
\ND\bigl((\tau w - \eparm_1)/\eparm_2,\, \eparm_2^{-2}\bigr)$, and $\tau$ is the 
standardized mean difference (Cohen's $d$); the mean difference on the original 
scale is $\tau/\eparm_2$. Relaxing $\h$ to a Bernstein polynomial removes the
normality assumption, recovering a flexible generalization of the Box-Cox model \citep{boxcox_1964}.

(ii) \textbf{Binary logistic regression and proportional odds models.} $G = \expit$ (the standard logistic distribution)
yields $\tau$ as a log-odds ratio. For a discrete, ordered outcome $\rY \in \{y_1
< \cdots < y_K\}$, parametrizing $\h$ as a step function, $\h(y_k) = \eparm_k$
for $k = 1, \ldots, K-1$ ($\eparm_K = \infty$), recovers binary logistic regression
when $K=2$ and the ordinal proportional odds model when $K>2$, under which the odds ratio
between treatment groups does not depend on which cutoff is used to dichotomize $\rY$.

(iii) \textbf{Cox and Weibull proportional hazards models.} $G = \cloglog^{-1}$ (the Gumbel
distribution) yields $\tau$ as a log-hazard ratio. For a time-to-event outcome 
$\rY \in \RR^+$, $\h$ is the log cumulative baseline hazard function; parametrizing
$\h$ as a Bernstein polynomial avoids imposing a distributional assumption on this
baseline hazard, giving a fully parametrized version of the Cox proportional
hazards model. Restricting $\h(\ry) = \eparm_1 + \eparm_2 \log(\ry)$ instead
gives the Weibull proportional hazards model.

\Needspace{5\baselineskip}
\subsection{Nonparanormal adjusted marginal inference}
\label{sec:adjusted}

The univariate transformation model of Section~\ref{sec:unadjusted} can be 
extended to leverage the prognostic information contained in baseline covariates 
$\rX$, while keeping the treatment effect $\tau$ marginal. This is achieved via 
NAMI, which jointly models $\rX$ and 
$\rY$ through a Gaussian copula \citep{Dandl_Hothorn_2026}, constructed such that
marginalizing the covariates out of the joint model recovers the same marginal 
treatment effect as in model~\eqref{eq:trafo-model}.

Fitting this joint model via a Gaussian copula requires transforming all marginal
distributions onto the standard normal scale. The marginal distribution of each
covariate $X_j$, $j = 1, \ldots, J-1$, is parametrized as an unconditional
transformation model,

\begin{equation}
    \Prob(X_j \le x_j) = \Phi\bigl(h_j(x_j)\bigr),
    \label{eq:marginalx}
\end{equation}

where the nondecreasing transformation function $h_j: \samX_j \to \RR$ transforms
$X_j$ to the latent standard normal scale. As for $\h$ in Section~\ref{sec:unadjusted},
$h_j$ can be parametrized flexibly, \eg as a step function for a discrete $X_j$ or
a Bernstein polynomial for a continuous one, without imposing a distributional
assumption on the covariates unless $h_j$ itself is restricted to a specific form.

The outcome $\rY$ is also transformed to the latent normal scale, via

\begin{equation}
    \h_J(\ry \mid w) := \Phi^{-1}\bigl[G\{\h(\ry) - \tau w\}\bigr],
    \label{eq:hJ}
\end{equation}

based on model~\eqref{eq:trafo-model}, where, as in Section~\ref{sec:unadjusted}, 
$G$ is chosen to give the desired treatment effect measure and $\h$ to adapt 
to the outcome type. If $G = \Phi$, this reduces to 
$\h_J(\ry \mid w) = \h(\ry) - \tau w$ directly.

Having transformed all marginals onto the standard normal scale, the joint
distribution of $\rX$ and $\rY$ can be modeled via a multivariate normal
distribution. Specifically, the multivariate transformation function
$\hvec: \samX_1 \times \cdots \times \samX_{J-1} \times \samY \to \RR^J$,

\[
\hvec(\rX, \rY \mid W = w) = \bigl(h_1(X_1), \ldots, h_{J-1}(X_{J-1}), \h_J(\rY \mid w)\bigr)^\top,
\]

defines the joint distribution of $\rX$ and $\rY$ given $W$ as

\begin{equation}
    \Prob(\rX \le \rx, \rY \le \ry \mid W = w) = \mPhi_{\mSigma}\bigl(\hvec(\rx, \ry \mid W = w)\bigr),
    \label{eq:joinedmultivar}
\end{equation}

where $\mPhi_{\mSigma}$ is the cumulative distribution function of
$\ND_J(\nullvec, \mSigma)$, the $J$-dimensional normal distribution with
correlation matrix $\mSigma$, parametrized via the inverse of its Cholesky
factor, $\mSigma = \mOmega^{-1}\mOmega^{-\top}$ \citep{Dandl_Hothorn_2026}.

Because $\mSigma$ has unit marginal variances,  marginalizing $\rX$ out of the 
joint model recovers exactly model~\eqref{eq:trafo-model} for $\rY$ given $W$:

\[
F_w(\ry) = \Phi\bigl(\h_J(\ry \mid w)\bigr)
= \Phi\Bigl[\Phi^{-1}\bigl\{G(\h(\ry) - \tau w)\bigr\}\Bigr]
= G\bigl(\h(\ry) - \tau w\bigr).
\]

The marginal interpretation of $\tau$ is therefore unchanged by covariate
adjustment.

From the joint model~\eqref{eq:joinedmultivar}, we can also derive the 
conditional distribution of $\rY$ given $W$ and $\rX$. For absolutely continuous 
variables, this gives:

\begin{equation}
    \Prob(\rY \le \ry \mid W = w, \rX = \rx)
    = \Phi\left(\sum_{j=1}^{J-1} \omega_{Jj} \h_j(x_j) + \omega_{JJ} \h_J(\ry \mid w)\right),
    \label{eq:condmultivar}
\end{equation}

with regression coefficients $\omega_{J1}, \ldots, \omega_{JJ}$ given by the $J$th
row of $\mOmega$. Because all covariates are transformed onto the same latent
normal scale, the coefficients $|\omega_{Jj}|$ are directly comparable and can
therefore be used to rank covariates by prognostic strength. The gain in
precision from adjusting for covariates can be quantified through the
coefficient of determination $R^2 := 1 - \omega_{JJ}^{-2}$.

Since $\h$ and each $h_j$ are fully parametrized, model~\eqref{eq:joinedmultivar}
is fully parametrized as well, and its parameters are estimated by likelihood
inference; details are given in \citep{hothorn_2024}.

\subsection{Randomization test on null-model score residuals}
\label{sec:likelihood_score}

Evidence for a marginal treatment effect in NAMI can be obtained using
likelihood-based procedures such as the Wald test, the score test, or the likelihood ratio
test. However, these methods rely on population-based inference, which assumes the
observed sample is a random sample from some larger population. We propose
instead to test the treatment effect using a permutation procedure, treating
the treatment assignment $W$, rather than the sample, as the random component.

Permutation tests of this kind can be built from a general linear statistic
\citep{Strasser1999:permutation,Hothorn2006:lego}, adapted here to test for
the independence of $W$ and $(\rX, \rY)$. For a binary treatment indicator,
$w_i \in \{0,1\}$, this takes the univariate form

\begin{equation}
    T = \sum_{i=1}^{N} w_i\, s(\rX_i, \rY_i) \in \RR,
    \label{eq:lego-teststat}
\end{equation}

where $s: \samX_1 \times \cdots \times \samX_{J-1} \times \samY \to \RR$ is
the score residual obtained from fitting NAMI under the null hypothesis
$H_0: \tau = 0$, which is permutation symmetric: its value does not depend
on the order in which the observations appear. In the case of more than
two treatment groups, the encoding of $W_i$ changes, and the test
statistic may become multivariate; see Section~\ref{sec:moretreats}.

Let $\ell_i(\thetavec, \lambdavec, \tau)$ denote the nonparanormal log-likelihood
for observation $i$ \citep{hothorn_2024}, where $\thetavec$
contains the parameters of $\h_1, \ldots, \h_{J-1}, \h_J$ and $\lambdavec$
the free lower-triangular parameters of $\mOmega$. Because $\mOmega$ is
constrained so that $\mSigma$ remains a valid correlation matrix, the
likelihood is parametrized in terms of $\lambdavec$ rather than $\mOmega$
directly \citep{hothorn_2024}.
Under the null model, we derive treatment-invariant score residuals by
replacing the treatment shift $\tau w$ in~\eqref{eq:hJ} with an auxiliary
location-shift parameter $\alpha$:

\begin{equation}
    \h_J(\ry) = \Phi^{-1}\bigl[G\bigl\{\h(\ry) - \alpha\bigr\}\bigr].
    \label{eq:perturbation}
\end{equation}

The resulting log-likelihood, $\ell_i(\thetavec, \lambdavec, \alpha)$, is used
to define the score residual at $\alpha = 0$ as

\begin{equation}
    s_i = \frac{\partial \ell_i(\thetavec, \lambdavec, \alpha)}{\partial \alpha}\bigg|_{\alpha = 0}.
    \label{eq:scoredef}
\end{equation}

For the special case where all $J$ variables are absolutely continuous and
observed without censoring, a closed-form derivation of $s_i$ is given in
Appendix~\ref{app:score}.

Under $H_0: \tau = 0$, the joint model is fitted to obtain estimates $\hat\thetavec$
and $\hat\lambdavec$, from which the scores $s_i$~\eqref{eq:scoredef}
and the test statistic $T$~\eqref{eq:lego-teststat} are computed. The
randomization distribution of the test statistic is obtained by holding
the observed pairs $(\rx_i, \ry_i)$ fixed and considering all permutations $\sigma \in \mathcal{S}$ of the
treatment labels $w_i$, where $\mathcal{S}$ is the set of $\binom{N}{N_1}$ possible
assignments and $N_1 = \sum_{i=1}^N w_i$ is the number of treated
observations.

The resulting $p$-value can be computed in three ways. The first, exact
enumeration, recomputes $T$ for every permutation $\sigma \in \mathcal{S}$ and reports
the proportion at least as extreme as the value observed; this becomes
computationally infeasible once $N$ is even moderately large. The
second, Monte Carlo resampling, approximates this proportion using a large
number of randomly drawn permutations instead of the full enumeration. The
third avoids resampling altogether by relying on the asymptotic distribution
of $T$ derived by \citet{Strasser1999:permutation}. 
Under $H_0$, given all permutations $\sigma \in \mathcal{S}$, the conditional
expectation and variance of $T$ are:

\begin{equation}
    \mathbb{E}(T \mid \mathcal{S}) = N_1 \bar{s}, \qquad
    \Var(T \mid \mathcal{S}) = \frac{N_1 N_0}{N - 1} \cdot
    \frac{1}{N}\sum_{i=1}^N \bigl(s_i - \bar{s}\bigr)^2,
    \label{eq:permmoments}
\end{equation}

where $\bar{s} = N^{-1}\sum_{i=1}^N s_i$ and $N_0 = N - N_1$ is the number of
control observations. Centering and scaling $T$ by these moments gives
$Z = \{T - \mathbb{E}(T \mid \mathcal{S})\}/\sqrt{\Var(T \mid \mathcal{S})}$, which is
asymptotically standard normal \citep{Strasser1999:permutation,Hothorn2006:lego}
and used to compute the $p$-value. Because $T$ is a scalar for the two treatment 
groups considered here, this probability is computed exactly. With more than two 
treatment groups, additional considerations may arise for computing the $p$-value, 
as discussed in Section~\ref{sec:moretreats}.

\subsection{Extensions}
\label{sec:extensions}

\subsubsection{Stratification}
\label{sec:stratification}

Under stratified randomization, observations belong to one of $B$ strata
(\eg study site), denoted $b_i \in \{1, \ldots, B\}$, and treatment is
randomized separately within each stratum. The outcome model can easily
be adapted to reflect this randomization structure by allowing each
stratum to have its own transformation function $\h_b: \samY \to \RR$
\citep{hothorn_moest_buehlmann_2017}, while keeping the treatment effect
$\tau$ constant across strata, so that model~\eqref{eq:trafo-model} becomes

\[
    F_w(\ry \mid b) = \Prob(\rY \le \ry \mid W = w, b)
    = G\bigl(\h_b(\ry) - \tau w\bigr).
\]

The corresponding transformation function used in the joint model,
analogous to~\eqref{eq:hJ}, is then:

\[
    \h_J(\ry \mid w, b) := \Phi^{-1}\bigl[G\{\h_b(\ry) - \tau w\}\bigr],
\]

with the treatment effect $\tau$ remaining common across strata, so that
only the distribution under control, not the treatment effect itself, is
allowed to differ between strata. The derivation of the score residual
then follows the same steps as in Section~\ref{sec:likelihood_score},
but now fitting the null model yields a separate estimate $\hat{\h}_b$
for each stratum.

The randomization test also needs to be adapted so that only within-strata
permutations are considered. The test statistic $T$, together with its
conditional expectation and variance, is computed separately within
each stratum and summed:

\[
T = \sum_{b=1}^{B} T_b, \qquad
\mathbb{E}(T \mid S) = \sum_{b=1}^{B} \mathbb{E}(T_b \mid S), \qquad
\Var(T \mid S) = \sum_{b=1}^{B} \Var(T_b \mid S).
\]

Although the value of $T$ is unchanged compared to the calculation across
all strata combined, its null distribution generally is not, and neither
is the resulting $p$-value. Asymptotically, the conditional expectation
and variance of $T$ change; for Monte Carlo or exact resampling, the set
of permutations considered changes.

\subsubsection{More than 2 treatment groups}
\label{sec:moretreats}

With $K > 2$ treatment groups, $W_i \in \{1, \ldots, K\}$, the binary
indicator $w_i$ in~\eqref{eq:lego-teststat} is no longer an adequate
encoding of treatment assignment, and an alternative encoding $a(W_i)$ is
required. We discuss two ways of choosing $a(W_i)$, depending on whether
the $K$ groups are ordered.

\textbf{Ordered treatment groups.} If it can reasonably be assumed that
the treatment effect varies monotonically with a natural
ordering of the $K$ groups, $1 < \cdots < K$ (\eg increasing dose
levels), $w_i$ can be replaced by a scalar encoding $a(W_i) =
u_{W_i}$, with $u_1 < \cdots < u_K$ reflecting this order. The simplest
choice is $u_k = k$, but $u_1, \ldots, u_K$ can instead be chosen to
reflect the magnitude of the difference between groups, for example using
the actual dose amounts \citep{Hothorn2006:lego}. The test statistic
remains univariate,
\[
T = \sum_{i=1}^N a(W_i)\, s_i,
\]
so the exact and Monte Carlo $p$-value calculations of
Section~\ref{sec:likelihood_score} still apply. Its asymptotic moments,
however, differ from~\eqref{eq:permmoments} and are instead given by:
\[
\begin{aligned}
\mathbb{E}(T\mid\mathcal S) &= \bar s \sum_{i=1}^N a(W_i), \\
\Var(T\mid\mathcal S) &= \frac{1}{N}\sum_{i=1}^N\bigl(s_i-\bar s\bigr)^2
\left[\frac{N}{N-1}\sum_{i=1}^N a(W_i)^2 - \frac{1}{N-1}\Bigl(\sum_{i=1}^N a(W_i)\Bigr)^{2}\right].
\end{aligned}
\]

\textbf{Unordered treatment groups.} Without a natural ordering, or when
monotonicity cannot reasonably be assumed, $a$
can instead be chosen as a dummy encoding of the $K$ groups, so that
$a(W_i) \in \mathbb{R}^{p \times 1}$ with $p = K$. Correspondingly, the
scalar treatment effect $\tau$ in~\eqref{eq:trafo-model} generalizes to a
$(K-1)$-dimensional vector, with one shift parameter for each group
relative to a reference group. In this case, the test
statistic generalizes to
\[
\boldsymbol T = \sum_{i=1}^N a(W_i)\, s_i \in \mathbb{R}^{p \times 1},
\]
which is multivariate \citep{Strasser1999:permutation,Hothorn2006:lego}
and must first be mapped into a single test statistic using its
conditional expectation $\boldsymbol\mu \in \mathbb{R}^{p \times 1}$ and
covariance $\boldsymbol\Sigma \in \mathbb{R}^{p \times p}$ under $H_0$.
Extending~\eqref{eq:permmoments} to the multivariate case:
\[
\begin{aligned}
\boldsymbol\mu &= \mathbb{E}(\boldsymbol T \mid \mathcal{S}) = \bar{s}\sum_{i=1}^N a(W_i), \\
\boldsymbol\Sigma &= \Var(\boldsymbol T \mid \mathcal{S}) \\
&= \frac{1}{N}\sum_{i=1}^N \bigl(s_i - \bar{s}\bigr)^2
\left[\frac{N}{N-1}\sum_{i=1}^N a(W_i)a(W_i)^\top -
\frac{1}{N-1}\Bigl(\sum_{i=1}^N a(W_i)\Bigr)\Bigl(\sum_{i=1}^N a(W_i)\Bigr)^\top\right].
\end{aligned}
\]
Two standard choices are the
maximum-type statistic $c_{\max}$ and the quadratic-form statistic
$c_{\mathrm{quad}}$,
\[
\begin{aligned}
c_{\max}(\boldsymbol T,\boldsymbol\mu,\boldsymbol\Sigma) &= \max \left|\frac{\boldsymbol T-\boldsymbol\mu}{\mathrm{diag}(\boldsymbol\Sigma)^{1/2}}\right|, \\
c_{\mathrm{quad}}(\boldsymbol T,\boldsymbol\mu,\boldsymbol\Sigma) &= (\boldsymbol T-\boldsymbol\mu)^\top \boldsymbol\Sigma^{+} (\boldsymbol T-\boldsymbol\mu),
\end{aligned}
\]
where $\boldsymbol\Sigma^{+}$ is the Moore-Penrose inverse of
$\boldsymbol\Sigma$. The choice of the test statistic depends on the
alternative of interest: $c_{\max}$ is most powerful when only one group
differs from the rest, while $c_{\mathrm{quad}}$ is sensitive to any kind
of differences between the groups.
As for $T$ in Section~\ref{sec:likelihood_score}, the $p$-value for
$c_{\max}$ or $c_{\mathrm{quad}}$ can be obtained by exact enumeration,
Monte Carlo resampling, or an asymptotic approximation. For the
asymptotic approximation, if $c_{\max}$ is chosen, the $p$-value is
obtained from the multivariate normal distribution of $\boldsymbol T$
and must be evaluated numerically; if $c_{\mathrm{quad}}$ is chosen,
this distribution simplifies instead to $\chi^2$ with $K-1$ degrees of
freedom \citep{Hothorn2006:lego}.
 
\section{Empirical evaluation}
\label{sec:simulation}

\citet{Dandl_Hothorn_2026} showed empirically that the Wald-based test of NAMI can
have inflated type~I error, especially as the number of covariates grows relative
to the sample size. We evaluate whether randomization-based inference for NAMI instead
maintains the nominal $5\%$ level, varying the prognostic strength of a covariate
and the number of noise covariates included. To do so, we rerun the simulation study
of \citet{Dandl_Hothorn_2026} and add NAMI with randomization-based inference to
the comparison.

\paragraph{Experimental setup}

The experiments of \citet{Dandl_Hothorn_2026} were based on a binomially distributed
treatment indicator $W \sim B(1, 0.5)$ and a potentially prognostic covariate
$X_1 \sim \chi^2_5$. Three outcome types were considered: a normally distributed
$\rY \sim \ND(w\tau, 1)$, for which $\tau$ reflects a Cohen's $d$; a binomially
distributed $\rY$, for which $\tau$ is a log-odds ratio; and a Weibull-distributed
$\rY$, for which $\tau$ is a log-hazard ratio, with heavy independent right-censoring
added. The prognostic strength of $X_1$ on $\rY$ was varied from absent to weak,
moderate, and strong. Additionally, $P-1$ correlated, $t$-distributed noise
covariates $X_2, \ldots, X_P$ were sampled independently of $X_1$ and $\rY$, for
$P \in \{1, 5, 15\}$ (including $X_1$). Sample sizes ($N = 82$ for continuous,
$N = 322$ for binary, and $N = 262$ for survival outcomes) were chosen by
\citet{Dandl_Hothorn_2026} to give approximately 60\% power to detect $\tau = 0.5$;
here, we instead focus only on $\tau = 0$ and report empirical size, based on
$10{,}000$ replications per scenario \citep[for further details, see]{Dandl_Hothorn_2026}.
Scripts reproducing all results in this paper are available in the
\proglang{R} package~\pkg{tram} (see Appendix~\ref{sec:software}).

We compared the empirical size of NAMI with randomization-based inference against
the methods studied by \citet{Dandl_Hothorn_2026}: NAMI with a Wald test, unadjusted
marginal inference (MI) with a Wald test, and a linear transformation model (LTM)
with a Wald test. We also report the 95\% Wald confidence interval of the empirical size
for each method.
Because the permutation-testing framework leaves the computation of the treatment effect 
unchanged, our point estimates closely match those reported by \citet{Dandl_Hothorn_2026} 
(see Appendix~\ref{app:simulation} for a comparison of NAMI estimates).
Small numerical differences are attributable to different software versions.
Unlike NAMI, LTM targets the conditional treatment effect directly and is therefore 
subject to noncollapsibility. MI, by contrast, ignores all covariates, so its 
results are unaffected by the number of covariates $P$ or the strength of the 
prognostic effect. We therefore report MI results only for $P = 1$ with no prognostic effect.

\paragraph{Results}

As shown in Table~\ref{tab:size}, NAMI with randomization-based inference maintained 
the nominal level of $5\%$ across all scenarios, regardless of the number of covariates or
prognostic strength. In contrast, the Wald tests for LTM and NAMI showed increasing size
distortion as the number of noise covariates grew, most notably for LTM with survival
outcomes and strong prognostic effects, where the empirical size exceeded $10\%$.

\begin{sidewaystable}[]
\centering
\caption{Empirical experiments: Empirical size and its 95\% Wald confidence interval
   based on $10,000$ replications 
  for different outcome types obtained from   unadjusted marginal inference (MI), 
  linear transformation models (LTM), and \NAMI (NAMI)
  under varying prognostic strength of covariate $X_1$ (in columns) and varying number of (noise) covariates ($P$, in rows).  \label{tab:size}} 
\begingroup\fontsize{9pt}{10pt}\selectfont
\begin{tabular}{llllllll}
  \hline & & & & \multicolumn{4}{c}{Empirical size} \\ 
\cmidrule(lr){5-8} 
Problem & Algorithm & Test & P & None & Weak & Moderate & Strong \\ 
  \hline
Continuous & MI & Wald & P = 1 & 0.053 (0.049, 0.058) &  &  &  \\ 
   & LTM & Wald & P = 1 & 0.058 (0.053, 0.062) & 0.055 (0.051, 0.06) & 0.054 (0.05, 0.059) & 0.051 (0.047, 0.056) \\ 
   &  &  & P = 5 & 0.059 (0.054, 0.064) & 0.062 (0.057, 0.067) & 0.062 (0.057, 0.066) & 0.06 (0.055, 0.065) \\ 
   &  &  & P = 15 & 0.083 (0.077, 0.088) & 0.083 (0.078, 0.089) & 0.079 (0.074, 0.085) & 0.084 (0.078, 0.089) \\ 
   & NAMI & Wald & P = 1 & 0.057 (0.053, 0.062) & 0.054 (0.05, 0.059) & 0.055 (0.05, 0.059) & 0.056 (0.052, 0.061) \\ 
   &  &  & P = 5 & 0.06 (0.056, 0.065) & 0.062 (0.057, 0.066) & 0.066 (0.061, 0.071) & 0.061 (0.057, 0.066) \\ 
   &  &  & P = 15 & 0.091 (0.085, 0.096) & 0.087 (0.081, 0.093) & 0.088 (0.083, 0.094) & 0.088 (0.083, 0.094) \\ 
   &  & Permutation & P = 1 & 0.05 (0.046, 0.054) & 0.048 (0.044, 0.052) & 0.049 (0.044, 0.053) & 0.048 (0.044, 0.053) \\ 
   &  &  & P = 5 & 0.048 (0.044, 0.053) & 0.049 (0.044, 0.053) & 0.052 (0.048, 0.056) & 0.048 (0.044, 0.052) \\ 
   &  &  & P = 15 & 0.051 (0.046, 0.055) & 0.05 (0.046, 0.054) & 0.051 (0.047, 0.055) & 0.051 (0.047, 0.055) \\ 
   \hline
Binary & MI & Wald & P = 1 & 0.049 (0.045, 0.053) &  &  &  \\ 
   & LTM & Wald & P = 1 & 0.049 (0.045, 0.053) & 0.051 (0.047, 0.055) & 0.05 (0.046, 0.054) & 0.055 (0.05, 0.059) \\ 
   &  &  & P = 5 & 0.051 (0.046, 0.055) & 0.052 (0.048, 0.056) & 0.051 (0.047, 0.055) & 0.052 (0.048, 0.056) \\ 
   &  &  & P = 15 & 0.059 (0.054, 0.063) & 0.052 (0.047, 0.056) & 0.056 (0.051, 0.06) & 0.061 (0.056, 0.066) \\ 
   & NAMI & Wald & P = 1 & 0.049 (0.044, 0.053) & 0.053 (0.048, 0.057) & 0.05 (0.046, 0.054) & 0.058 (0.053, 0.063) \\ 
   &  &  & P = 5 & 0.051 (0.047, 0.056) & 0.052 (0.047, 0.056) & 0.052 (0.048, 0.056) & 0.052 (0.048, 0.056) \\ 
   &  &  & P = 15 & 0.06 (0.055, 0.064) & 0.054 (0.05, 0.058) & 0.057 (0.052, 0.061) & 0.065 (0.06, 0.069) \\ 
   &  & Permutation & P = 1 & 0.049 (0.044, 0.053) & 0.052 (0.048, 0.057) & 0.049 (0.045, 0.054) & 0.056 (0.051, 0.06) \\ 
   &  &  & P = 5 & 0.05 (0.045, 0.054) & 0.05 (0.046, 0.054) & 0.05 (0.045, 0.054) & 0.047 (0.043, 0.051) \\ 
   &  &  & P = 15 & 0.054 (0.05, 0.059) & 0.048 (0.044, 0.052) & 0.051 (0.046, 0.055) & 0.049 (0.045, 0.054) \\ 
   \hline
Survival & MI & Wald & P = 1 & 0.052 (0.047, 0.056) &  &  &  \\ 
   & LTM & Wald & P = 1 & 0.052 (0.047, 0.056) & 0.055 (0.051, 0.06) & 0.066 (0.061, 0.071) & 0.104 (0.098, 0.11) \\ 
   &  &  & P = 5 & 0.054 (0.049, 0.058) & 0.053 (0.048, 0.057) & 0.066 (0.061, 0.07) & 0.101 (0.095, 0.107) \\ 
   &  &  & P = 15 & 0.064 (0.059, 0.069) & 0.066 (0.061, 0.071) & 0.085 (0.079, 0.09) & 0.111 (0.105, 0.117) \\ 
   & NAMI & Wald & P = 1 & 0.053 (0.048, 0.057) & 0.053 (0.049, 0.058) & 0.054 (0.05, 0.058) & 0.06 (0.055, 0.065) \\ 
   &  &  & P = 5 & 0.055 (0.05, 0.059) & 0.052 (0.047, 0.056) & 0.056 (0.051, 0.06) & 0.056 (0.051, 0.06) \\ 
   &  &  & P = 15 & 0.059 (0.055, 0.064) & 0.064 (0.059, 0.069) & 0.069 (0.064, 0.074) & 0.071 (0.066, 0.076) \\ 
   &  & Permutation & P = 1 & 0.05 (0.045, 0.054) & 0.051 (0.047, 0.055) & 0.051 (0.046, 0.055) & 0.054 (0.049, 0.058) \\ 
   &  &  & P = 5 & 0.049 (0.045, 0.053) & 0.046 (0.042, 0.05) & 0.049 (0.045, 0.053) & 0.046 (0.042, 0.05) \\ 
   &  &  & P = 15 & 0.047 (0.043, 0.052) & 0.051 (0.046, 0.055) & 0.054 (0.05, 0.058) & 0.05 (0.046, 0.054) \\ 
   \hline
\end{tabular}
\endgroup
\end{sidewaystable}

\section{Application to 20 randomized controlled trials}
\label{sec:application}

We apply the proposed method to 20 phase~III RCTs from a pharmaceutical
company, conducted between 2006 and 2023.
Identifying details are not reported to preserve confidentiality.
For each trial, patients were randomized to two treatment groups with a 1:1 or 2:1 allocation
ratio. All trials used stratified randomization,
with the number of stratification factors ranging from \numberstringnum{1} to
\numberstringnum{4}. Sample sizes ranged from around $150$ to $1500$ patients, with an
interquartile range (IQR) of 324--666. Between \numberstringnum{1} and \numberstringnum{5}
baseline covariates were available per trial for adjustment. Table~\ref{tab:characteristics}
in Appendix~\ref{app:trial-characteristics} gives these characteristics for each trial individually.

For each trial, we focus on a time-to-event endpoint and fit three models:
(i) a stratified Cox model without covariate adjustment, (ii) NAMI with
$G = \cloglog^{-1}$ and a Bernstein polynomial baseline transformation (Section~\ref{sec:unadjusted}),
and (iii) the covariate-adjusted log-rank approach of \citet{Ye_Shao_Yi_2024}, under
an assumed Cox model, which we abbreviate as LRCL. Computational details are given
in Appendix~\ref{sec:software}. All three models stratify on the same
stratification factors, with NAMI and LRCL additionally adjusting for the same baseline
covariates. Each model therefore estimates a log-hazard ratio, conditional on the
stratification factors but marginal with respect to the covariates, assuming proportional
hazards and independent right-censoring, without imposing a distributional assumption
on the baseline hazard. For each model, we extract the log-hazard ratio and its
standard error to compute a two-sided Wald $p$-value; for NAMI, we additionally compute
a two-sided asymptotic permutation $p$-value (Section~\ref{sec:stratification}). We also extract NAMI's $R^2$
(Section~\ref{sec:adjusted}), which quantifies the share of outcome variance explained
by the covariates.

Covariate adjustment is expected to reduce the standard error of the treatment effect
estimate, provided the covariates are prognostic. We quantify this efficiency gain
as the percentage reduction in the number of events needed to reach the same
standard error as the unadjusted stratified Cox model,
$100 \times \{1 - (\mathrm{SE}_{\text{adj}}/\mathrm{SE}_{\text{unadj}})^2\}$.
Figure~\ref{fig:fig_pct-events-reduction-dotplot} shows this reduction for
NAMI and LRCL, ordered by increasing $R^2$; Table~\ref{tab:per-study} in
Appendix~\ref{app:results} gives more detailed results for all three models.

\begin{Schunk}
\begin{figure}[ht]

{\centering \includegraphics[width=1\linewidth]{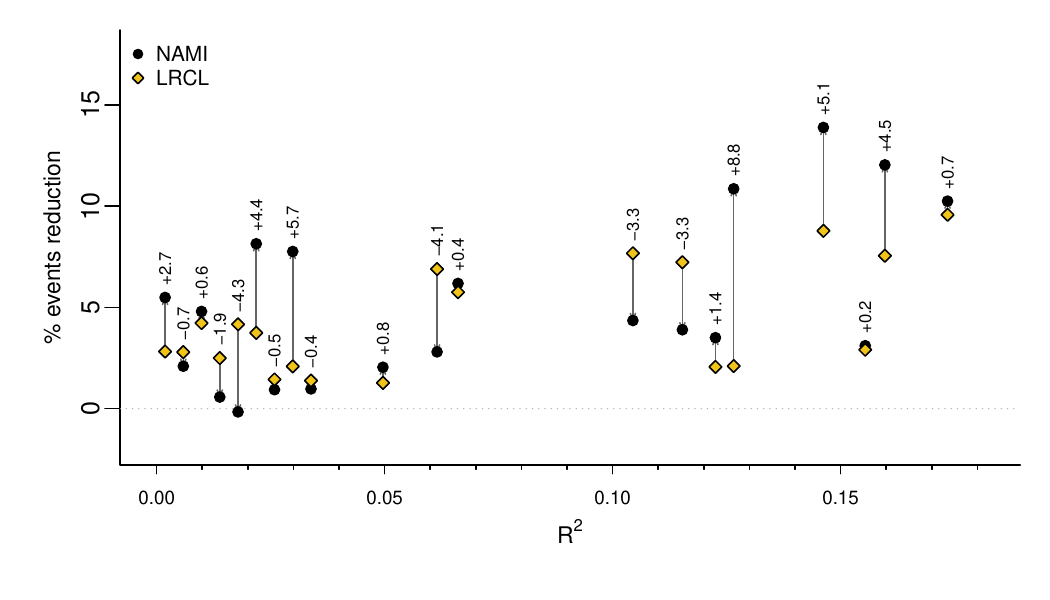} 

}

\caption[Percentage reduction in events achieved by covariate adjustment for NAMI and LRCL, plotted against $R^2$]{Percentage reduction in events achieved by covariate adjustment for NAMI and LRCL, plotted against $R^2$. Each point shows $100\times\{1-(\mathrm{SE}_{\text{adj}}/\mathrm{SE}_{\text{unadj}})^2\}$, where $\mathrm{SE}_{\text{adj}}$ is the standard error of the corresponding adjusted model (NAMI or LRCL) and $\mathrm{SE}_{\text{unadj}}$ is the standard error of the unadjusted stratified Cox model. Note that $R^2$ is only defined for NAMI, so both methods are plotted against NAMI's $R^2$ for a given trial. For visual clarity, a horizontal jitter was added for trials with near-identical $R^2$ values. Vertical lines connect the two estimates within each trial, and the number above each pair gives the percentage-point difference (NAMI minus LRCL), with positive values indicating a larger efficiency gain for NAMI.}\label{fig:fig_pct-events-reduction-dotplot}
\end{figure}
\end{Schunk}

Overall, the efficiency gained from covariate adjustment was modest: a median
reduction in events of 4.1\% for NAMI (IQR: 2.1--7.9\%)
and 3.3\% for LRCL (IQR: 2.1--7\%).
This is consistent with NAMI's $R^2$, which was small in most trials (median:
0.06, IQR: 0.02--0.12), meaning the covariates
were generally not strongly prognostic. Trials with higher $R^2$ tended to have
larger gains in efficiency, and $R^2$ itself tended to be somewhat larger in the
smaller trials (median of 0.12 among the 5 smallest trials
versus 0.03 among the 5 largest). Still, across
all 20 trials, this reduction in standard error never changed which trials showed
evidence of a treatment effect at the 5\% level, for either NAMI or LRCL, compared
to the unadjusted stratified Cox model (Table~\ref{tab:per-study}).
One trial showed a slightly negative percentage reduction for NAMI (-0.2\%, $R^2 = 0.016$), which is possible since NAMI has no guarantee against small 
efficiency losses when covariates carry little to no prognostic information.
The results between NAMI and LRCL were otherwise similar, with neither method 
producing systematically larger reductions in standard error or smaller $p$-values 
than the other: NAMI had larger reductions in standard error in 12
of 20 trials and smaller Wald $p$-values in 10.

\begin{Schunk}
\begin{figure}[ht]

{\centering \includegraphics[width=1\linewidth]{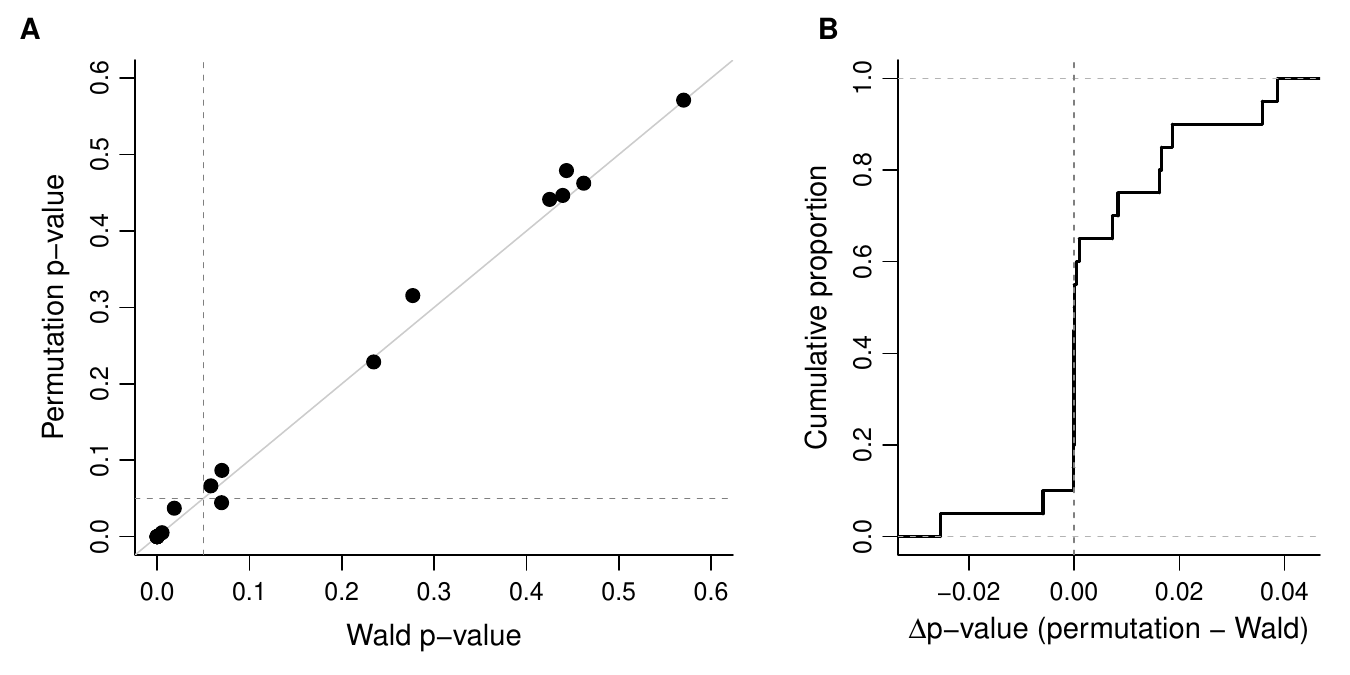} 

}

\caption[Comparison of the Wald and permutation $p$-values of NAMI]{Comparison of the Wald and permutation $p$-values of NAMI. Panel A shows a scatterplot of the two $p$-values for each trial, with the diagonal line indicating equality and dashed lines denoting the 5\% significance threshold on each axis. Panel B shows the empirical cumulative distribution function of the difference between the two $p$-values (permutation minus Wald): the x-axis gives the $p$-value difference and the y-axis gives the cumulative proportion of trials at or below that difference, with the dashed vertical line at zero.}\label{fig:fig_NAMI-pvals-composite-scatter}
\end{figure}
\end{Schunk}

Figure~\ref{fig:fig_NAMI-pvals-composite-scatter} compares the Wald and permutation
$p$-values of NAMI across the 20 trials. The two agree
closely, differing by a median absolute difference of 0.001 (up to
0.039 at most), with a tendency for the permutation $p$-value to be
larger than the Wald $p$-value (14 of 20 trials).
Only one trial showed disagreement between the two methods on whether there was
evidence of a treatment effect at the 5\% level, where the Wald $p$-value was
0.07 and the permutation $p$-value was 0.044, both
close to the significance threshold. Given the large sample sizes and the small number
of covariates used across these trials, this level of agreement is what would generally
be expected.
 
\section{Discussion}
\label{sec:discussion}

Following Fisher's argument that randomization provides a physical basis for inference
\citep{Fisher1935:design}, we proposed a randomization-based test for NAMI.
NAMI adjusts for prognostic baseline covariates while preserving the marginal
interpretation of the treatment effect, including for noncollapsible effect
measures such as log-odds and log-hazard ratios.
Rather than relying on population-based inference, the proposed test derives
its reference distribution from the known treatment randomization mechanism.
This allows covariate-adjusted inference without requiring patients to be
assumed to be a random sample from a larger population. The framework accommodates
different outcome types and censoring structures and can be adapted to
stratified randomization designs.

Our simulation studies confirmed that the proposed test maintained the nominal
$5\%$ level across continuous, binary, and time-to-event outcomes, including
settings where the Wald test for NAMI showed increasing type~I error inflation
as the number of covariates grew. This illustrates the potential benefit of
randomization-based inference, particularly in trials with small sample sizes
or many covariates relative to the number of observations. Consistent with
this, the Wald and permutation $p$-values agreed closely in the application to
20 historical randomized trials, where sample sizes were large and the number
of covariates was small. The estimate from NAMI also generally gained
precision from the prognostic information in the covariates, with gains
similar to those of the stratified covariate-adjusted log-rank test
\citep{Ye_Shao_Yi_2024}.

A practical limitation of randomization-based NAMI is computational cost. The randomization test requires
fitting NAMI under the null hypothesis, which becomes demanding as the number
of covariates grows, since each covariate contributes its own marginal
parameters as well as a correlation parameter with every other variable and
the outcome. Once this fit is obtained, however, the score residuals can be
reused across treatment assignments, so no refitting is required for
individual permutations. 

Furthermore, the randomization-based framework assumes a known randomization
mechanism, so that the permutation scheme can mirror the actual allocation. We
focused on randomized controlled trials with simple or stratified
randomization, where permuting treatment labels suffices. For more complex
designs, such as adaptive randomization, deriving the reference distribution is
more intricate. For observational data, or for trials in which censoring or
dropout depends on treatment, exchanging treatment labels is difficult to
justify at all.

A further direction is the development of randomization-based confidence
intervals for NAMI. In principle, such intervals could be constructed by
inverting a sequence of randomization tests at different candidate values of
the treatment effect.
However, test inversion would require
reevaluating the test statistic across candidate values, which would increase
computational costs. Interpolation-based search strategies could
reduce this cost by avoiding an exhaustive grid of candidate values, though
whether this preserves the coverage properties of the resulting interval
remains to be investigated.
 
\paragraph{Acknowledgements}

Susanne Dandl was financially supported by the Swiss National 
Science Foundation under grant number 
200021\_219384. Leticia W\"uthrich was supported 
by a PhD fellowship funded by Novartis AG. We thank Novartis AG for
providing access to the trial data used to evaluate the application of
this method.

\bibliography{NAMI,packages}

\newpage

\renewcommand\thefigure{S.~\arabic{figure}}    
\renewcommand\thetable{S.~\arabic{table}}
\renewcommand\thesection{A.\arabic{section}}
\setcounter{figure}{0}
\setcounter{table}{0}
\setcounter{section}{0}

\appendix


\section{Derivation of the score residual}
\label{app:score}

Assuming all $J$ variables are absolutely continuous and observed without
censoring, the joint log-likelihood for observation $i$ based on the flow
nonparanormal approximation \citep{hothorn_2024} is

\begin{eqnarray*}
    \ell_i(\thetavec, \lambdavec, \tau) &=& \log\Bigl(\text{\boldmath$\phi$}\bigl(\bigl(\h_1(x_{i1}), \ldots, \h_{J-1}(x_{i,J-1}),
    \h_J(\ry_i \mid w_i)\bigr)^\top \mid \mOmega\bigr)\Bigr) \nonumber \\
    && + \sum_{j=1}^{J-1} \log\bigl(\h_j'(x_{ij})\bigr) + \log\bigl(\h_J'(\ry_i \mid w_i)\bigr),
\end{eqnarray*}

with $\h_J(\ry \mid w) = \Phi^{-1}[G\{\h(\ry) - \tau w\}]$ as in~\eqref{eq:hJ}.
We derive the score residual $s_i$ under $H_0: \tau = 0$, replacing $\tau w$
with the auxiliary location-shift parameter $\alpha$, which gives

\begin{eqnarray*}
    \ell_i(\thetavec, \lambdavec, \alpha) &=& \log\Bigl(\text{\boldmath$\phi$}\bigl(\bigl(\h_1(x_{i1}), \ldots, \h_{J-1}(x_{i,J-1}),
    \h_J(\ry_i)\bigr)^\top \mid \mOmega\bigr)\Bigr) \nonumber \\
    && + \sum_{j=1}^{J-1} \log\bigl(\h_j'(x_{ij})\bigr) + \log\bigl(\h_J'(\ry_i)\bigr).
\end{eqnarray*}

The terms $\log\bigl(\h_j'(x_{ij})\bigr)$, $j = 1, \ldots,
J-1$, do not contain $\alpha$ at all, so they vanish when we differentiate
wrt $\alpha$ and can be dropped.

By the definition of the $J$-variate normal density parameterized through
the lower triangular $\mOmega$ \citep{hothorn_2024},

\[
\log\Bigl(\text{\boldmath$\phi$}\bigl(\zvec_i \mid \mOmega\bigr)\Bigr)
= -\frac{J}{2}\log(2\pi) + \sum_{k=1}^J \log(\omega_{kk})
- \frac{1}{2} \bigl(\lVert \mOmega \zvec_i \rVert_2\bigr)^{2},
\]

where

\[
\zvec_i = \bigl(\h_1(x_{i1}), \ldots, \h_{J-1}(x_{i,J-1}), \h_J(\ry_i)\bigr)^\top.
\]

The first two terms are constants that do not depend on $\alpha$ and
can be dropped as well.
Writing out the Euclidean norm as a sum over $k$, and each entry of
$\mOmega \zvec_i$ as a sum over $j$, where $\omega_{kj}$ denotes the
$(k,j)$th entry of $\mOmega$,

\[
\bigl(\lVert \mOmega \zvec_i \rVert_2\bigr)^{2} = \sum_{k=1}^J (\mOmega \zvec_i)_k^2
= \sum_{k=1}^J \left( \sum_{j=1}^{J-1} \omega_{kj}\, \h_j(x_{ij})
+ \omega_{kJ}\, \h_J(\ry_i) \right)^2.
\]

Therefore, the log-likelihood kernel can be written as

\[
\tilde\ell_i(\alpha) =
\underbrace{-\frac{1}{2}\sum_{k=1}^{J}
\left( \sum_{j=1}^{J-1} \omega_{kj}\, \h_j(x_{ij}) + \omega_{kJ}\, \h_J(\ry_i) \right)^{2}}_{A}
\;+\; \underbrace{\log\bigl(\h_J'(\ry_i)\bigr)}_{B}.
\]

The derivatives of $A$ and $B$ are obtained separately below and combined
to give $s_i = \partial \tilde\ell_i(\alpha)/\partial\alpha \,\big|_{\alpha = 0}$.

\subsection*{Derivative of $A$}

Write $A = -\tfrac12\sum_{k=1}^J u_k(\alpha)^2$ with
$u_k(\alpha) = \sum_{j=1}^{J-1}\omega_{kj}\h_j(x_{ij}) + \omega_{kJ}\h_J(\ry_i)$.
Only the last term of $u_k$ depends on $\alpha$, so by the chain rule,

\[
\frac{\partial A}{\partial \alpha}
= -\sum_{k=1}^J u_k(\alpha) \cdot \omega_{kJ} \cdot \frac{\partial \h_J(\ry_i)}{\partial \alpha}
= -\frac{\partial \h_J(\ry_i)}{\partial \alpha}
\sum_{k=1}^J \omega_{kJ}\, u_k(\alpha).
\]

Because $\mOmega$ is lower triangular, $\omega_{kJ} = 0$ for all $k < J$, so
only $k = J$ contributes:

\[
\frac{\partial A}{\partial \alpha}
= -\frac{\partial \h_J(\ry_i)}{\partial \alpha} \cdot
\omega_{JJ}\left( \sum_{j=1}^{J-1}\omega_{Jj}\,\h_j(x_{ij}) + \omega_{JJ}\,\h_J(\ry_i) \right).
\]

$\partial \h_J(\ry)/\partial \alpha$ can be found using the rule
for differentiating an inverse function, $(\Phi^{-1})'(v) =
1/\phi(\Phi^{-1}(v))$, together with the chain rule,

\[
\frac{\partial \h_J(\ry)}{\partial \alpha}
= \underbrace{\bigl(\Phi^{-1}\bigr)'\bigl(G\{\h(\ry)-\alpha\}\bigr)}_{=\ 1/\phi(\h_J(\ry))}
\cdot\ g\bigl(\h(\ry) - \alpha\bigr) \cdot (-1)
= \frac{-\, g\bigl(\h(\ry) - \alpha\bigr)}{\phi\bigl(\h_J(\ry)\bigr)},
\]

so that, evaluating $\h_J(\ry)$ at $\alpha = 0$ gives $\Phi^{-1}[G\{\h(\ry)\}]$,

\[
\frac{\partial \h_J(\ry)}{\partial \alpha}\bigg|_{\alpha = 0}
= \frac{-\, g\bigl(\h(\ry)\bigr)}{\phi\bigl(\Phi^{-1}[G\{\h(\ry)\}]\bigr)}
= -\, \frac{g\bigl(\h(\ry)\bigr)}{\phi\bigl(\h_J(\ry)\bigr)}.
\]

Substituting this into the expression for $\partial A/\partial\alpha$ above,
evaluated at $\alpha = 0$,

\[
\frac{\partial A}{\partial \alpha}\bigg|_{\alpha = 0}
= \frac{g\bigl(\h(\ry_i)\bigr)}{\phi\bigl(\h_J(\ry_i)\bigr)} \cdot
\omega_{JJ}\left( \sum_{j=1}^{J-1}\omega_{Jj}\,\h_j(x_{ij}) + \omega_{JJ}\,\h_J(\ry_i) \right).
\]

\subsection*{Derivative of $B$}

$B = \log\bigl(\h_J'(\ry_i)\bigr)$ requires
$\h_J'(\ry) := \partial \h_J(\ry)/\partial \ry$
first. As in Derivative of $A$, but now wrt $\ry$ instead of $\alpha$, the
same rule for differentiating an inverse function together with the chain
rule gives

\[
\h_J'(\ry)
= \underbrace{\bigl(\Phi^{-1}\bigr)'\bigl(G\{\h(\ry)-\alpha\}\bigr)}_{=\ 1/\phi(\h_J(\ry))}
\cdot\ g\bigl(\h(\ry) - \alpha\bigr) \cdot \h'(\ry)
= \frac{g\bigl(\h(\ry) - \alpha\bigr)\, \h'(\ry)}{\phi\bigl(\h_J(\ry)\bigr)}.
\]

Taking logs,

\[
\log\bigl(\h_J'(\ry)\bigr) = \log\Bigl(g\bigl(\h(\ry) - \alpha\bigr)\Bigr) + \log\bigl(\h'(\ry)\bigr)
- \log\Bigl(\phi\bigl(\h_J(\ry)\bigr)\Bigr).
\]

Differentiating wrt $\alpha$, the middle term,
$\log\bigl(\h'(\ry)\bigr)$, does not depend on $\alpha$. For the first term,

\[
\frac{\partial}{\partial \alpha} \log\Bigl(g\bigl(\h(\ry) - \alpha\bigr)\Bigr)
= \frac{g'\bigl(\h(\ry) - \alpha\bigr)}{g\bigl(\h(\ry) - \alpha\bigr)} \cdot (-1),
\]

so that

\[
\frac{\partial}{\partial \alpha} \log\Bigl(g\bigl(\h(\ry) - \alpha\bigr)\Bigr)\bigg|_{\alpha = 0}
= -\frac{g'\bigl(\h(\ry)\bigr)}{g\bigl(\h(\ry)\bigr)}.
\]

For the third term, using $\phi'(z) = -z\,\phi(z)$ so that
$\phi'(z)/\phi(z) = -z$, together with $\partial \h_J(\ry)/\partial\alpha$
from Derivative of $A$,

\[
\frac{\partial}{\partial \alpha} \log\Bigl(\phi\bigl(\h_J(\ry)\bigr)\Bigr)
= \frac{\phi'\bigl(\h_J(\ry)\bigr)}{\phi\bigl(\h_J(\ry)\bigr)}
\cdot \frac{\partial \h_J(\ry)}{\partial \alpha}
= -\h_J(\ry) \cdot \frac{\partial \h_J(\ry)}{\partial \alpha},
\]

so that

\[
\frac{\partial}{\partial \alpha} \log\Bigl(\phi\bigl(\h_J(\ry)\bigr)\Bigr)\bigg|_{\alpha = 0}
= \h_J(\ry)\cdot\frac{g(\h(\ry))}{\phi(\h_J(\ry))}.
\]

The derivative of $B$ at $\alpha = 0$ is the first term minus the third
term:

\[
\frac{\partial B}{\partial \alpha}\bigg|_{\alpha = 0}
= -\frac{g'\bigl(\h(\ry_i)\bigr)}{g\bigl(\h(\ry_i)\bigr)}
\;-\; \h_J(\ry_i) \cdot \frac{g\bigl(\h(\ry_i)\bigr)}{\phi\bigl(\h_J(\ry_i)\bigr)}.
\]

\subsection*{Combining the derivatives}

Adding the two derivatives, and writing $g = G'$ and $\varrho_i := g\bigl(\h(\ry_i)\bigr) /
\phi\bigl(\h_J(\ry_i)\bigr)$, gives the score residual for observation $i$:

\begin{equation}
    s_i
    = \varrho_i \cdot \omega_{JJ}
    \left( \sum_{j=1}^{J-1} \omega_{Jj}\, \h_j(x_{ij})
    + \omega_{JJ}\, \h_J(\ry_i) \right)
    \;-\; \frac{g'\bigl(\h(\ry_i)\bigr)}{g\bigl(\h(\ry_i)\bigr)}
    \;-\; \h_J(\ry_i) \cdot \varrho_i.
    \label{eq:score}
\end{equation}

In the special case $G = \Phi$, we have $g = \phi$, $\h_J(\ry_i) = \h(\ry_i)$,
and $\varrho_i = 1$, so the score reduces to

\begin{equation}
    s_i
    = \omega_{JJ}
    \left( \sum_{j=1}^{J-1} \omega_{Jj}\, \h_j(x_{ij})
    + \omega_{JJ}\, \h_J(\ry_i) \right).
    \label{eq:score-normal}
\end{equation}

\clearpage
\section{Empirical evaluation}
\label{app:simulation}

\begin{figure}[ht!]
\centering
\begin{knitrout}
\definecolor{shadecolor}{rgb}{0.969, 0.969, 0.969}\color{fgcolor}
\includegraphics[width=.9\textwidth]{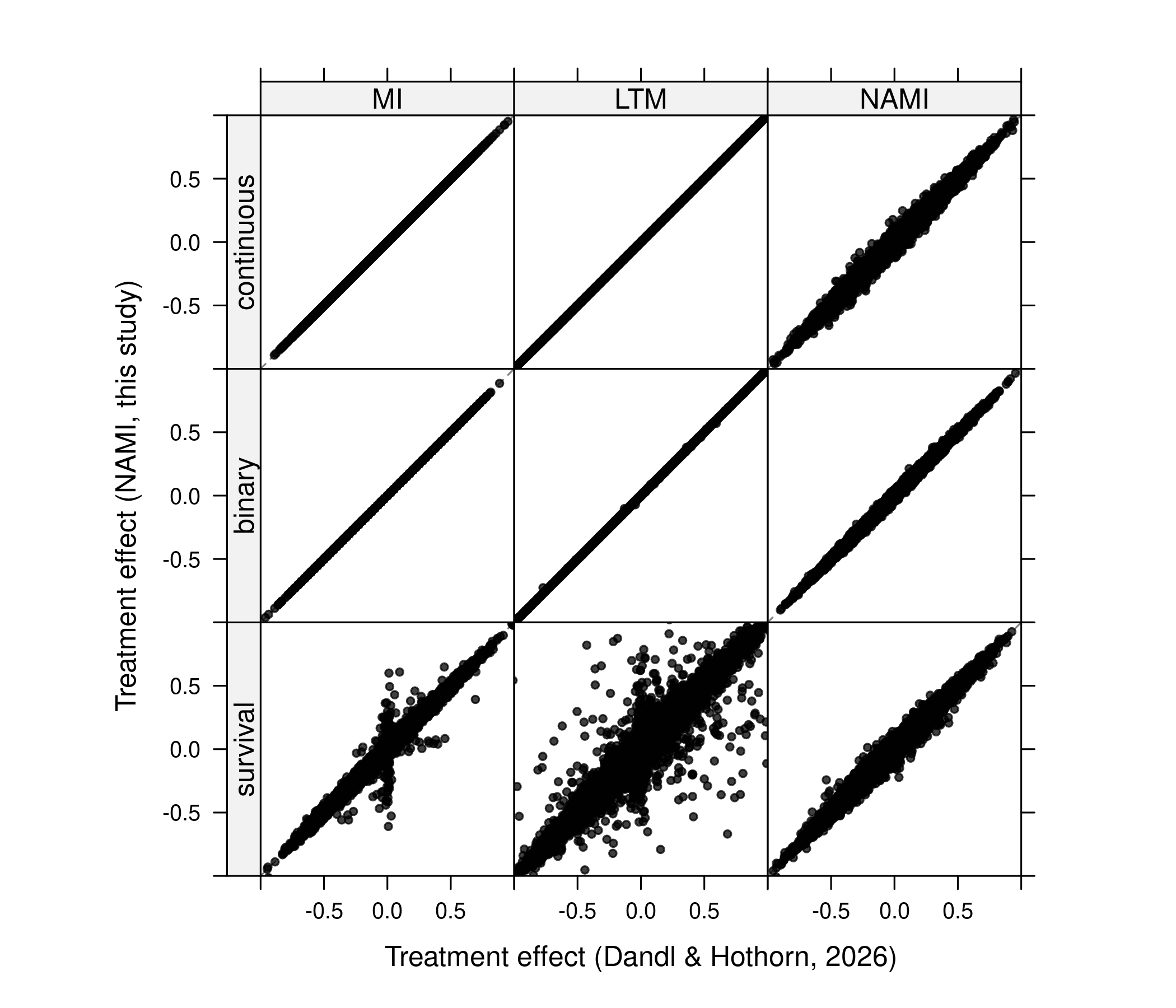} 
\end{knitrout}
\caption{Treatment effect estimates obtained in this study versus those
reported by \citet{Dandl_Hothorn_2026} for different outcome types and
methods: unadjusted marginal inference (MI), \NAMI (NAMI), and linear
transformation models (LTM).
The dashed line indicates identical estimates. \label{fig:comparison}}
\end{figure}

\clearpage
\section{Application to 20 randomized controlled trials}
\label{app:application}

\subsection{Trial characteristics}
\label{app:trial-characteristics}

\begin{table}[H]
\centering
\caption{Characteristics of the 20 trials, ordered by increasing sample size. Columns give sample size (binned into four categories), the number of stratification factors used for randomization and analysis, and the number of baseline covariates used for adjustment, split into discrete and continuous counts.}
\label{tab:characteristics}
\small
\begin{tabular}{c c c ccc}
\toprule
& & & \multicolumn{3}{c}{\textbf{Covariates}} \\
\cmidrule(lr){4-6}
\textbf{Trial} & \textbf{Sample size} & \textbf{Strata} & \textbf{Total} & \textbf{Discrete} & \textbf{Continuous} \\
\midrule
1 & 150-300 & 1 & 2 & 1 & 1 \\
2 & 150-300 & 2 & 3 & 1 & 2 \\
3 & 150-300 & 2 & 3 & 2 & 1 \\
4 & 150-300 & 2 & 3 & 2 & 1 \\
5 & 300-500 & 1 & 5 & 2 & 3 \\
6 & 300-500 & 1 & 3 & 2 & 1 \\
7 & 300-500 & 2 & 4 & 2 & 2 \\
8 & 300-500 & 2 & 5 & 1 & 4 \\
9 & 300-500 & 1 & 2 & 1 & 1 \\
10 & 300-500 & 2 & 1 & 0 & 1 \\
11 & 500-700 & 3 & 1 & 0 & 1 \\
12 & 500-700 & 1 & 4 & 2 & 2 \\
13 & 500-700 & 3 & 2 & 1 & 1 \\
14 & 500-700 & 3 & 4 & 2 & 2 \\
15 & 500-700 & 2 & 2 & 1 & 1 \\
16 & 500-700 & 3 & 2 & 1 & 1 \\
17 & 700-1500 & 2 & 4 & 3 & 1 \\
18 & 700-1500 & 3 & 3 & 2 & 1 \\
19 & 700-1500 & 4 & 2 & 0 & 2 \\
20 & 700-1500 & 3 & 1 & 0 & 1 \\
\bottomrule
\end{tabular}
\end{table}

\subsection{Results}
\label{app:results}

Table~\ref{tab:per-study} reports results per trial. To avoid identifying individual
trials, only the p-value is given for the unadjusted stratified Cox model; for NAMI and LRCL,
the log-hazard ratio and standard error are reported relative to the unadjusted model.

\small
\setlength{\tabcolsep}{4pt}
\begin{longtable}{c l c c c c c}
\caption{\normalsize Per-trial results for the unadjusted stratified Cox model, NAMI, and LRCL, ordered by increasing sample size. Only the Wald p-value is reported for the unadjusted stratified Cox model. For NAMI and LRCL, the difference in log-hazard ratio and the percentage reduction in events are reported relative to the unadjusted stratified Cox model, along with the Wald p-value; $R^2$ and the randomization p-value are additionally reported for NAMI.}
\label{tab:per-study}\\
\toprule
 & & & & & \multicolumn{2}{c}{\textbf{p-value}} \\
\cmidrule(lr){6-7}
 \textbf{Trial} & \textbf{Method} & \textbf{$\Delta$logHR} & \textbf{\% reduction} & \textbf{$R^2$} & \textbf{Wald} & \textbf{Perm.} \\
 \midrule
\endfirsthead
\toprule
 & & & & & \multicolumn{2}{c}{\textbf{p-value}} \\
\cmidrule(lr){6-7}
 \textbf{Trial} & \textbf{Method} & \textbf{$\Delta$logHR} & \textbf{\% reduction} & \textbf{$R^2$} & \textbf{Wald} & \textbf{Perm.} \\
 \midrule
\endhead
\midrule
\multicolumn{7}{r}{\small\itshape continued on next page}\\
\endfoot
\bottomrule
\endlastfoot
                    & Cox      & --    & --  & --   & $<$ 0.0001 & --  \\
1 & NAMI     & -0.04 & 6.2 & 0.066 & $<$ 0.0001 & $<$ 0.0001 \\
                    & LRCL     & 0.01 & 5.7 & --   & $<$ 0.0001 & --  \\
\addlinespace
\rowcolor{gray!6}
                    & \rule{0pt}{2.6ex}Cox      & --    & --  & --   & $<$ 0.0001 & --  \\
\rowcolor{gray!6}
2 & NAMI     & 0.08 & 3.9 & 0.12 & $<$ 0.0001 & $<$ 0.0001 \\
\rowcolor{gray!6}
                    & \rule[-1.1ex]{0pt}{0pt}LRCL     & 0.06 & 7.2 & --   & $<$ 0.0001 & --  \\
\addlinespace
                    & Cox      & --    & --  & --   & 0.011 & --  \\
3 & NAMI     & 0.09 & 10.9 & 0.13 & 0.019 & 0.037 \\
                    & LRCL     & -0.02 & 2.1 & --   & 0.009 & --  \\
\addlinespace
\rowcolor{gray!6}
                    & \rule{0pt}{2.6ex}Cox      & --    & --  & --   & 0.39 & --  \\
\rowcolor{gray!6}
4 & NAMI     & -0.02 & 12.0 & 0.16 & 0.43 & 0.44 \\
\rowcolor{gray!6}
                    & \rule[-1.1ex]{0pt}{0pt}LRCL     & 0.00 & 7.5 & --   & 0.36 & --  \\
\addlinespace
                    & Cox      & --    & --  & --   & 0.15 & --  \\
5 & NAMI     & 0.11 & 4.4 & 0.10 & 0.44 & 0.45 \\
                    & LRCL     & 0.12 & 7.7 & --   & 0.46 & --  \\
\addlinespace
\rowcolor{gray!6}
                    & \rule{0pt}{2.6ex}Cox      & --    & --  & --   & $<$ 0.0001 & --  \\
\rowcolor{gray!6}
6 & NAMI     & 0.00 & -0.2 & 0.016 & $<$ 0.0001 & $<$ 0.0001 \\
\rowcolor{gray!6}
                    & \rule[-1.1ex]{0pt}{0pt}LRCL     & 0.03 & 4.2 & --   & $<$ 0.0001 & --  \\
\addlinespace
                    & Cox      & --    & --  & --   & 0.075 & --  \\
7 & NAMI     & -0.00 & 3.1 & 0.16 & 0.07 & 0.087 \\
                    & LRCL     & -0.01 & 2.9 & --   & 0.064 & --  \\
\addlinespace
\rowcolor{gray!6}
                    & \rule{0pt}{2.6ex}Cox      & --    & --  & --   & 0.087 & --  \\
\rowcolor{gray!6}
8 & NAMI     & -0.00 & 10.2 & 0.17 & 0.07 & 0.044 \\
\rowcolor{gray!6}
                    & \rule[-1.1ex]{0pt}{0pt}LRCL     & 0.05 & 9.6 & --   & 0.15 & --  \\
\addlinespace
                    & Cox      & --    & --  & --   & 0.32 & --  \\
9 & NAMI     & 0.02 & 2.1 & 0.011 & 0.28 & 0.32 \\
                    & LRCL     & 0.04 & 2.8 & --   & 0.25 & --  \\
\addlinespace
\rowcolor{gray!6}
                    & \rule{0pt}{2.6ex}Cox      & --    & --  & --   & $<$ 0.0001 & --  \\
\rowcolor{gray!6}
10 & NAMI     & -0.02 & 0.6 & 0.015 & $<$ 0.0001 & $<$ 0.0001 \\
\rowcolor{gray!6}
                    & \rule[-1.1ex]{0pt}{0pt}LRCL     & -0.01 & 2.5 & --   & $<$ 0.0001 & --  \\
\addlinespace
                    & Cox      & --    & --  & --   & 0.058 & --  \\
11 & NAMI     & 0.00 & 8.1 & 0.017 & 0.058 & 0.067 \\
                    & LRCL     & 0.02 & 3.7 & --   & 0.086 & --  \\
\addlinespace
\rowcolor{gray!6}
                    & \rule{0pt}{2.6ex}Cox      & --    & --  & --   & 0.22 & --  \\
\rowcolor{gray!6}
12 & NAMI     & 0.03 & 3.5 & 0.12 & 0.23 & 0.23 \\
\rowcolor{gray!6}
                    & \rule[-1.1ex]{0pt}{0pt}LRCL     & 0.02 & 2.1 & --   & 0.23 & --  \\
\addlinespace
                    & Cox      & --    & --  & --   & 0.42 & --  \\
13 & NAMI     & -0.00 & 5.5 & 0.009 & 0.44 & 0.48 \\
                    & LRCL     & -0.03 & 2.8 & --   & 0.56 & --  \\
\addlinespace
\rowcolor{gray!6}
                    & \rule{0pt}{2.6ex}Cox      & --    & --  & --   & 0.25 & --  \\
\rowcolor{gray!6}
14 & NAMI     & 0.06 & 13.9 & 0.15 & 0.46 & 0.46 \\
\rowcolor{gray!6}
                    & \rule[-1.1ex]{0pt}{0pt}LRCL     & 0.07 & 8.8 & --   & 0.55 & --  \\
\addlinespace
                    & Cox      & --    & --  & --   & 0.008 & --  \\
15 & NAMI     & -0.03 & 0.9 & 0.023 & 0.005 & 0.005 \\
                    & LRCL     & -0.01 & 1.4 & --   & 0.007 & --  \\
\addlinespace
\rowcolor{gray!6}
                    & \rule{0pt}{2.6ex}Cox      & --    & --  & --   & $<$ 0.0001 & --  \\
\rowcolor{gray!6}
16 & NAMI     & -0.02 & 1.0 & 0.027 & $<$ 0.0001 & $<$ 0.0001 \\
\rowcolor{gray!6}
                    & \rule[-1.1ex]{0pt}{0pt}LRCL     & -0.01 & 1.4 & --   & $<$ 0.0001 & --  \\
\addlinespace
                    & Cox      & --    & --  & --   & $<$ 0.0001 & --  \\
17 & NAMI     & -0.02 & 2.8 & 0.061 & $<$ 0.0001 & $<$ 0.0001 \\
                    & LRCL     & -0.01 & 6.9 & --   & $<$ 0.0001 & --  \\
\addlinespace
\rowcolor{gray!6}
                    & \rule{0pt}{2.6ex}Cox      & --    & --  & --   & $<$ 0.0001 & --  \\
\rowcolor{gray!6}
18 & NAMI     & -0.01 & 4.8 & 0.015 & $<$ 0.0001 & $<$ 0.0001 \\
\rowcolor{gray!6}
                    & \rule[-1.1ex]{0pt}{0pt}LRCL     & 0.00 & 4.2 & --   & $<$ 0.0001 & --  \\
\addlinespace
                    & Cox      & --    & --  & --   & $<$ 0.0001 & --  \\
19 & NAMI     & 0.01 & 7.8 & 0.027 & $<$ 0.0001 & $<$ 0.0001 \\
                    & LRCL     & -0.03 & 2.1 & --   & $<$ 0.0001 & --  \\
\addlinespace
\rowcolor{gray!6}
                    & \rule{0pt}{2.6ex}Cox      & --    & --  & --   & 0.58 & --  \\
\rowcolor{gray!6}
20 & NAMI     & -0.00 & 2.0 & 0.05 & 0.57 & 0.57 \\
\rowcolor{gray!6}
                    & \rule[-1.1ex]{0pt}{0pt}LRCL     & -0.06 & 1.3 & --   & 0.37 & --  \\
\addlinespace
\end{longtable}
\normalsize
\setlength{\tabcolsep}{6pt}

\section{Computational details}
\label{sec:software}

All computations were performed using \proglang{R} \citep{R}. 
The simulation study in Section~\ref{sec:simulation} and the application 
to 20 phase~III RCTs in Section~\ref{sec:application} fit NAMI
using the \pkg{tram} and \pkg{mlt} packages \citep{pkg:tram,pkg:mlt}. 
The source code to reproduce the simulation study is available in the 
\pkg{tram} package. 
After installing the package, the location of the code can be found with

\begin{knitrout}
\definecolor{shadecolor}{rgb}{0.969, 0.969, 0.969}\color{fgcolor}\begin{kframe}
\begin{alltt}
\hlkwd{system.file}\hldef{(}\hlsng{"nami_simulations"}\hldef{,} \hlkwc{package} \hldef{=} \hlsng{"tram"}\hldef{)}
\end{alltt}
\end{kframe}
\end{knitrout}

In the application, the unadjusted stratified Cox model was fit with the
\pkg{survival} package \citep{pkg:survival}, and LRCL with the
\pkg{RobinCar} package \citep{pkg:RobinCar}. The permutation $p$-value was
computed using the \pkg{coin} package \citep{pkg:coin}. Within NAMI,
discrete baseline covariates were modeled using \texttt{tram::Polr}
with \texttt{method = "probit"}, continuous baseline covariates using
\texttt{tram::BoxCox}, and the outcome using a stratified \texttt{tram::Coxph}
model.

Due to data sharing privacy concerns, the datasets from the application section cannot
be shared. We instead apply the analysis code run on the datasets on a publicly
available time-to-event dataset:
the \pkg{CAOsurv} data from the German CAO/ARO/AIO-04 rectal cancer trial
\citep{roedel_graeven_fietkau_2015}, available via the
\pkg{TH.data} package. The dataset includes 1236 patients, with
623 randomized to fluorouracil-based
chemoradiotherapy alone and 613 to
fluorouracil plus oxaliplatin, with disease-free survival (\code{iDFS}) as the outcome
and randomization stratified by two factors (\code{strat\_t}, \code{strat\_n}). We
adjust for age, BMI, sex, and ECOG performance status as baseline covariates.

We first load the required packages and the data.

\begin{knitrout}
\definecolor{shadecolor}{rgb}{0.969, 0.969, 0.969}\color{fgcolor}\begin{kframe}
\begin{alltt}
\hlkwd{library}\hldef{(}\hlsng{"tram"}\hldef{)}      \hlcom{# to fit marginal and joint models}
\hlkwd{library}\hldef{(}\hlsng{"coin"}\hldef{)}      \hlcom{# to do permutation test}
\hlkwd{library}\hldef{(}\hlsng{"TH.data"}\hldef{)}
\hlkwd{load}\hldef{(}\hlkwd{system.file}\hldef{(}\hlsng{"rda"}\hldef{,} \hlsng{"Primary_endpoint_data.rda"}\hldef{,} \hlkwc{package} \hldef{=} \hlsng{"TH.data"}\hldef{))}
\end{alltt}
\end{kframe}
\end{knitrout}

The stratification variable and a selection of baseline covariates are then prepared.

\begin{knitrout}
\definecolor{shadecolor}{rgb}{0.969, 0.969, 0.969}\color{fgcolor}\begin{kframe}
\begin{alltt}
\hldef{CAOsurv}\hlopt{$}\hldef{strat} \hlkwb{<-} \hlkwd{with}\hldef{(CAOsurv,} \hlkwd{interaction}\hldef{(strat_t, strat_n))}
\hldef{CAOsurv}\hlopt{$}\hldef{sex}   \hlkwb{<-} \hlkwd{factor}\hldef{(CAOsurv}\hlopt{$}\hldef{geschlecht,}
                        \hlkwc{levels} \hldef{=} \hlkwd{levels}\hldef{(CAOsurv}\hlopt{$}\hldef{geschlecht))}
\hldef{CAOsurv}\hlopt{$}\hldef{ecog}  \hlkwb{<-} \hlkwd{factor}\hldef{(CAOsurv}\hlopt{$}\hldef{ecog_b,} \hlkwc{ordered} \hldef{=} \hlnum{TRUE}\hldef{)}
\hldef{CAOsurv}\hlopt{$}\hldef{BMI}   \hlkwb{<-} \hldef{CAOsurv}\hlopt{$}\hldef{gewicht}\hlopt{/}\hldef{(CAOsurv}\hlopt{$}\hldef{groesse}\hlopt{/}\hlnum{100}\hldef{)}\hlopt{^}\hlnum{2}
\end{alltt}
\end{kframe}
\end{knitrout}

Each covariate is modeled unconditionally, and the outcome is modeled conditional on
treatment, stratified by the randomization factors.

\begin{knitrout}
\definecolor{shadecolor}{rgb}{0.969, 0.969, 0.969}\color{fgcolor}\begin{kframe}
\begin{alltt}
\hlcom{# Marginal covariate models:}
\hldef{m_age}  \hlkwb{<-} \hldef{tram}\hlopt{::}\hlkwd{BoxCox}\hldef{(age} \hlopt{~} \hlnum{1}\hldef{,} \hlkwc{data} \hldef{= CAOsurv,} \hlkwc{na.action} \hldef{= na.pass)}
\hldef{m_BMI}  \hlkwb{<-} \hldef{tram}\hlopt{::}\hlkwd{BoxCox}\hldef{(BMI} \hlopt{~} \hlnum{1}\hldef{,} \hlkwc{data} \hldef{= CAOsurv,} \hlkwc{na.action} \hldef{= na.pass)}
\hldef{m_sex}  \hlkwb{<-} \hldef{tram}\hlopt{::}\hlkwd{Polr}\hldef{(sex} \hlopt{~} \hlnum{1}\hldef{,} \hlkwc{method} \hldef{=} \hlsng{"probit"}\hldef{,} \hlkwc{data} \hldef{= CAOsurv,}
                     \hlkwc{na.action} \hldef{= na.pass)}
\hldef{m_ecog} \hlkwb{<-} \hldef{tram}\hlopt{::}\hlkwd{Polr}\hldef{(ecog} \hlopt{~} \hlnum{1}\hldef{,} \hlkwc{method} \hldef{=} \hlsng{"probit"}\hldef{,} \hlkwc{data} \hldef{= CAOsurv,}
                     \hlkwc{na.action} \hldef{= na.pass)}
\hlcom{# Marginal outcome model:}
\hldef{m_Y}    \hlkwb{<-} \hldef{tram}\hlopt{::}\hlkwd{Coxph}\hldef{(iDFS} \hlopt{|} \hldef{strat} \hlopt{~} \hldef{randarm,} \hlkwc{data} \hldef{= CAOsurv,}
                      \hlkwc{na.action} \hldef{= na.pass)}
\end{alltt}
\end{kframe}
\end{knitrout}

These marginal models are combined into the joint NAMI model, fitted under the null
hypothesis of no treatment effect.

\begin{knitrout}
\definecolor{shadecolor}{rgb}{0.969, 0.969, 0.969}\color{fgcolor}\begin{kframe}
\begin{alltt}
\hldef{m0} \hlkwb{<-} \hldef{mlt}\hlopt{::}\hlkwd{mmlt}\hldef{(m_age, m_BMI, m_sex, m_ecog,} \hlcom{# <-- first, covariate models}
                \hldef{m_Y,}                         \hlcom{# <-- last, outcome model}
                \hlkwc{formula} \hldef{=} \hlopt{~}\hlnum{1}\hldef{,} \hlkwc{data} \hldef{= CAOsurv,}
                \hlkwc{fixed} \hldef{=} \hlkwd{c}\hldef{(}\hlsng{"iDFS.randarm5-FU + Oxaliplatin"} \hldef{=} \hlnum{0}\hldef{),}
                \hlkwc{args} \hldef{=} \hlkwd{list}\hldef{(}\hlkwc{type} \hldef{=} \hlsng{"ghalton"}\hldef{,} \hlkwc{M} \hldef{=} \hlnum{1000}\hldef{))}
\end{alltt}
\end{kframe}
\end{knitrout}

Finally, the score residuals are extracted from the null model and used to construct
the stratified randomization test.

\begin{knitrout}
\definecolor{shadecolor}{rgb}{0.969, 0.969, 0.969}\color{fgcolor}\begin{kframe}
\begin{alltt}
\hldef{r0} \hlkwb{<-} \hlkwd{residuals}\hldef{(m0)[,} \hlsng{"iDFS"}\hldef{]}
\hldef{coin}\hlopt{::}\hlkwd{independence_test}\hldef{(r0} \hlopt{~} \hldef{randarm} \hlopt{|} \hldef{strat,} \hlkwc{data} \hldef{= CAOsurv)}
\end{alltt}
\begin{verbatim}
## 
## 	Asymptotic General Independence Test
## 
## data:  r0 by
## 	 randarm (5-FU, 5-FU + Oxaliplatin) 
## 	 stratified by strat
## Z = -1.9163, p-value = 0.05532
## alternative hypothesis: two.sided
\end{verbatim}
\end{kframe}
\end{knitrout}
 
\end{document}